\documentclass[prl,aps,twocolumn,superscriptaddress]{revtex4}
\usepackage[]{graphicx}
\usepackage[]{hyperref}
\usepackage[]{amsmath}
\usepackage{braket}
\usepackage{color}
\usepackage{textcomp}

\begin{document}

\newcommand{\TwoEightSi}{\ensuremath{^{28}\text{Si}}}
\newcommand{\natSi}{\ensuremath{^{\text{nat}}\text{Si}}}

\newcommand{\TConcentration}{ $>1.7\times10^{13}$~cm$^{-3}$}

\newcommand{\chaiLWhot}{$329.8\pm0.6$~MHz}
\newcommand{\chaiSDhot}{$250\pm100$~MHz}

\newcommand{\aaSD}{$1.3\pm0.3$~GHz}
\newcommand{\aaSDbound}{$2.9\pm0.2$~GHz}
\newcommand{\aaLW}{$22.8\pm0.5$~GHz}

\newcommand{\qqSD}{$1.0\pm0.2$~GHz}
\newcommand{\qqSDbound}{$2.4\pm0.2$~GHz}
\newcommand{\qqLW}{$20.6\pm0.3$~GHz}

\newcommand{\taurusLW}{$6.0\pm0.2$~GHz}
\newcommand{\taurusLWnoPM}{$6.0$~GHz}
\newcommand{\taurusSD}{$16\pm7$~MHz}

\newcommand{\chaiSD}{$27\pm12$~MHz}
\newcommand{\chaiLW}{$38\pm6$~MHz}
\newcommand{\chaiLWnoPM}{$38$~MHz}

\newcommand{\taurusHotLW}{$6.67\pm0.09$~GHz}
\newcommand{\fzLW}{$9.01\pm0.08$~GHz}

\newcommand{\pZfit}{$P_0=46\pm3$~MHz}
\newcommand{\pTfit}{$P_{\text{T}}=35\pm8$~GHz}
\newcommand{\eAfit}{$E_{\text{a}}=1800\pm80$~\textmu eV}
\newcommand{\thermSD}{$2.0\pm0.4$~MHz}

\title{$T$ centres in photonic silicon-on-insulator material}

\date{\today}
\author{E. R. MacQuarrie}
\affiliation{Department of Physics, Simon Fraser University, Burnaby, British Columbia, Canada}
\affiliation{Photonic Inc., Vancouver, British Columbia, Canada}

\author{C. Chartrand}
\affiliation{Department of Physics, Simon Fraser University, Burnaby, British Columbia, Canada}

\author{D. B. Higginbottom}
\affiliation{Department of Physics, Simon Fraser University, Burnaby, British Columbia, Canada}

\author{K. J. Morse}
\affiliation{Department of Physics, Simon Fraser University, Burnaby, British Columbia, Canada}
\affiliation{Photonic Inc., Vancouver, British Columbia, Canada}

\author{V. A. Karasyuk}
\affiliation{Department of Physics, Simon Fraser University, Burnaby, British Columbia, Canada}

\author{S. Roorda}
\affiliation{Department of Physics, Universit\'{e} de Montr\'{e}al, Montreal, Quebec, Canada}

\author{S. Simmons}
\affiliation{Department of Physics, Simon Fraser University, Burnaby, British Columbia, Canada}
\affiliation{Photonic Inc., Vancouver, British Columbia, Canada}
\email{s.simmons@sfu.ca}

\begin{abstract}

Global quantum networks will benefit from the reliable creation and control of high-performance solid-state telecom photon-spin interfaces. \textit{T} radiation damage centres in silicon provide a promising photon-spin interface due to their narrow $O$-band optical transition near $1326$~nm and long-lived electron and nuclear spin lifetimes. To date, these defect centres have only been studied as ensembles in bulk silicon. Here, we demonstrate the reliable creation of high concentration \textit{T} centre ensembles in the 220~nm device layer of silicon-on-insulator (SOI) wafers by ion implantation and subsequent annealing. We then develop a method that uses spin-dependent optical transitions to benchmark the characteristic optical spectral diffusion within these \textit{T} centre ensembles. Using this new technique, we show that with minimal optimization to the fabrication process high densities of implanted \textit{T} centres localized $\lesssim100$~nm from an interface display $\sim1$~GHz characteristic levels of total spectral diffusion. 

\end{abstract}

\maketitle

\section{Introduction}

Point defect colour centres in solid state systems provide a promising platform for emerging quantum technologies. 
A number of colour centres are currently in development~\cite{awschalom2018, zhang2020}, but few of these leading candidates are both natively integrated into silicon and intrinsically operate at telecom wavelengths. 
Radiation damage centres in silicon have recently emerged as promising light-matter interfaces that simultaneously meet both of these criteria~\cite{buckley2017, chartrand2018, beaufils2018, redjem2020}. 
Within this family of defects, the \textit{T} centre notably possesses highly coherent electron and nuclear spin degrees of freedom and narrow, spin-dependent ensemble optical transitions near $1326$~nm in the telecommunications $O$-band~\cite{bergeron2020}.
These properties make the \textit{T} centre a competitive candidate for integration into future commercial-scale quantum networks with long-lived quantum memory and computing capabilities.

Incorporating \textit{T} centres into such quantum networks requires a reliable means of making the centres in device-ready material such as silicon-on-insulator (SOI). 
Other photon-spin interfaces have been produced by ion implantation and thermal treatment~\cite{meijer2005, chu2011, falk2013, beaufils2018, phenicie2019, buckley2020, redjem2020}. 
Here, we use similar techniques to generate high concentrations of \textit{T} centres in the 220nm device layer of commercial SOI wafers. 
We achieve \textit{T} centre concentrations $>30$ times higher than in previously measured bulk samples and, in follow-on work \cite{kurkjian2021}, present evidence that these SOI concentrations are~\TConcentration. Moreover, the inhomogeneous broadening of the zero phonon line (ZPL) within implanted ensembles is only $\sim35\%$ larger than in \textit{T} centre ensembles generated by electron irradiating and annealing comparable bulk material. 

The fabrication process used to generate \textit{T} centres should preserve the excellent spin and optical properties that \textit{T} centre ensembles exhibit in bulk silicon where the majority of centres are located far from material interfaces.
The increase in electric and magnetic field noise near those interfaces can dramatically degrade a centre's spin and optical properties. 
Encouragingly, donor spin defects in silicon have already demonstrated competitive electron and nuclear spin coherence times ($>0.5$~s, $>30$~s) at distances of just $10$~nm from \TwoEightSi /SiO$_2$ interfaces \cite{muhonen2014}, and \textit{T} centres can be expected to demonstrate comparable or better spin properties at our target depth of $110$~nm. 
In the optical domain, spectral diffusion --- the time-varying changes in a centre's optical transition energies --- often dominates the optical properties of emitters located near interfaces.
Spectral diffusion can reduce the indistinguishability of emitted photons, limiting an emitter's potential in quantum technologies. 
Measuring the characteristic spectral diffusion of the generated \textit{T} centre ensembles thus provides an important metric for benchmarking our fabrication process. 
To this end, we develop a method that uses spin-dependent optical transitions to efficiently measure the characteristic long-term (total) spectral diffusion within an ensemble of emitters. 

Applying this new technique to our high-concentration \textit{T} centre ensembles in SOI, we measure characteristic total spectral diffusion of $\sim1$~GHz. Although this value is $\sim10^4$ times larger than the $169$~kHz lifetime-limited linewidth, excellent photon indistinguishability could still be achieved by employing fabrication or measurement techniques designed to stabilize the electromagnetic environment~\cite{oliveira2017, anderson2019, sangtawesin2019, kasperczyk2020}, by filtering the emitted photons~\cite{bernien2013, gao2019}, or by Purcell-enhancing the radiative decay rate~\cite{grange2015, giesz2015} as the \textit{T} centre is expected to have a high radiative efficiency~\cite{bergeron2020}.

\section{T Centre Implantation}

\begin{figure}[ht]
\includegraphics[width=\linewidth]{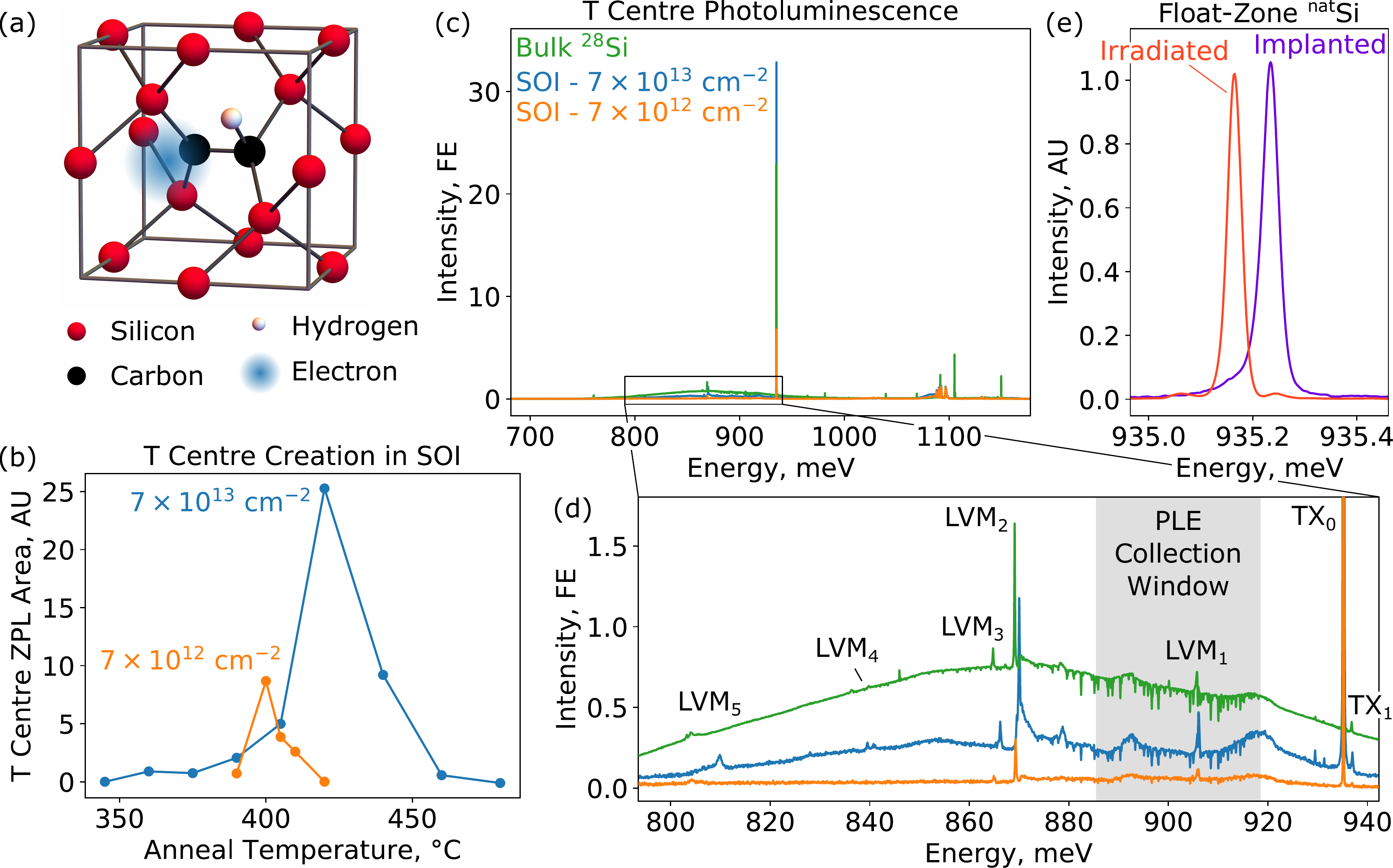} \\
\caption[fig:fig1] {\textit{T} centre implantation development. (a) Atomic structure of the \textit{T} centre in Si as proposed in Ref.~\cite{safanov1996}. (b) Integrated \textit{T} centre zero phonon line (ZPL) area in SOI as a function of the final anneal temperature. (c,d) \textit{T} centre photoluminescence (PL) at a temperature of $4.2$~K for bulk $^{28}$Si and SOI implanted at fluences of $7\times10^{12}$~cm$^{-2}$ and $7\times10^{13}$~cm$^{-2}$. The measured signal has been normalized to the intensity of the free exciton (FE) recombination line. Panel (d) highlights the collection window for our photoluminescence excitation (PLE) spectroscopy measurements. The sharp dips within this window are water vapor absorption lines. (e) ZPL PL for \textit{T} centres in bulk float-zone $^{\text{nat}}$Si that were created either by implantation or electron irradiation. The signal has been normalized to the fitted \textit{T} centre ZPL amplitude. }
\label{fig:fig1}
\end{figure}

As shown in Fig.~\ref{fig:fig1}a, the \textit{T} centre is thought to consist of two carbon atoms sharing the substitutional site of one silicon atom; an interstitial hydrogen terminates one of these carbon atoms while the other remains stable with an unpaired electron~\cite{safanov1996}. The current formation model for this defect consists of an interstitial carbon capturing a hydrogen atom and then migrating to a substitutional carbon site during heat treatment at temperatures between $350^{\circ}$ and $600^{\circ}$C~\cite{safanov1996, bergeron2020}. Starting from SOI with a $220$~nm thick Czochralski (CZ) natural Si ($^{\text{nat}}$Si) device layer and $3$~\textmu m of buried SiO$_{2}$, we realize this formation process by implanting carbon and then hydrogen in a fixed [C:H]=1:1 dose ratio, annealing between the two implants and then again after the hydrogen implant. 

We performed two fabrication runs: one with a $^{12}$C implant and doses of $7\times10^{12}$~cm$^{-2}$ and a second with $^{13}$C and doses of $7\times10^{13}$~cm$^{-2}$. These doses, which were chosen from measurements detailed in the Supplementary Materials (SM)~\cite{SI}, place the initial carbon concentrations well above the carbon solubility limit for silicon at $1000^{\circ}$C~\cite{pichler2004}. Carbon implants were performed at an energy of $38$~keV, while hydrogen was implanted at $9$~keV to target overlapping implantation profiles with a mean depth of $110$~nm. 

In between the two implants, we perform a rapid thermal anneal at $1000^{\circ}$C for $20$~seconds in an argon background to repair lattice damage generated during the carbon implant and to incorporate carbon onto substitutional lattice sites~\cite{berhanuddin2012, beaufils2018}. The subsequent hydrogen implant then generates lattice damage and interstitial carbon to promote \textit{T} centre formation, but the smaller mass of the hydrogen ions generates less total damage than the carbon implant~\cite{SI}. After the hydrogen implant, we boil the samples for $1$~hour in deionized water to further increase the hydrogen concentration before performing a final rapid thermal anneal for $3$~min at a temperature ranging from $345^{\circ}$ to $480^{\circ}$C in a nitrogen atmosphere. 

Fig.~\ref{fig:fig1}b shows the integrated area of the \textit{T} centre ZPL as a function of the final anneal temperature for a series of SOI samples. For these samples, we find optimum temperatures of $400^{\circ}$C for a fluence of $7\times10^{12}$~cm$^{-2}$ and $420^{\circ}$C for $7\times10^{13}$~cm$^{-2}$. Unless otherwise noted, we have normalized photoluminescence (PL) spectra to the free exciton (FE) recombination line at $\sim1097$~meV. Although the FE recombination line can vary in intensity across samples as, for instance, the density of recombination sites changes, it provides a means of roughly quantifying the \textit{T} centre concentration to enable comparison amongst samples. 

In Fig.~\ref{fig:fig1}c,d, we plot the PL spectra under above-band ($532$~nm) excitation for these particular SOI samples alongside that of an isotopically-purified $^{28}$Si bulk sample with an initial $1.5\times10^{15}$~cm$^{-3}$ carbon concentration. \textit{T} centres were created in this bulk sample by electron irradiating and annealing (details in Ref.~\cite{bergeron2020}). Fig.~\ref{fig:fig1}d provides a detailed view of the \textit{T} centre sideband spectra, showing an isotopic shift in the local vibrational modes (LVMs) as the $7\times10^{13}$~cm$^{-2}$ SOI sample was implanted with $^{13}$C isotopes, the $7\times10^{12}$~cm$^{-2}$ SOI sample was implanted with $^{12}$C, and the bulk $^{28}$Si sample contains a natural distribution of carbon isotopes ($98.9\%$~$^{12}$C). In Fig.~\ref{fig:fig1}d, we have also labeled both ZPLs within the \textit{T} centre exciton doublet, where $\text{TX}_1$ sits $1.76$~meV above $\text{TX}_0$~\cite{bergeron2020}. 

By normalizing the integrated area of the \textit{T} centre ZPL to the nominal sample thickness, we estimate that the implanted \textit{T} centre densities for the $7\times10^{12}$~cm$^{-2}$ and $7\times10^{13}$~cm$^{-2}$ SOI samples are $>5$ and $>30$ times larger than the \textit{T} centre density in the bulk $^{28}$Si sample, respectively. Here, we have used $220$~nm as the SOI thickness and the $2.5$~\textmu m absorption depth of $532$~nm light in silicon at cryogenic temperatures~\cite{dash1955, jellison2018} as the nominal $^{28}$Si thickness. As the free excitons generated by the $532$~nm excitation are not confined to the absorption region in our bulk sample, this latter thickness serves as a lower bound, making our relative concentration estimates lower bounds on the true ratios. In the SM, we verify that the implanted \textit{T} centres are localized to the SOI device layer by selectively etching away the device layer and also by using ultraviolet PL excitation, which has a much shallower absorption depth than $532$~nm~\cite{SI}. We also note that this estimate of the \textit{T} centre density neglects differences in the defects' local optical environment that could modify their emission properties~\cite{barnes1998}. 

In Fig.~\ref{fig:fig1}e, we compare the ZPL PL for \textit{T} centres generated by implantation to that of \textit{T} centres created by electron irradiation. Both samples in this figure are bulk, float-zone (FZ) $^{\text{nat}}$Si. One was implanted with $^{13}$C and hydrogen at a fluence of $7\times10^{12}$~cm$^{-2}$, while the other was electron irradiated and annealed to convert native carbon and hydrogen into \textit{T} centres~\cite{bergeron2020}. We once again see the expected $^{13}$C isotopic shift in the implanted ZPL, and at $4.2$~K, we report ZPL linewidths of \taurusHotLW~in the irradiated FZ sample and \fzLW~for the implanted FZ wafer. As will be shown below, these linewidths lie well above the linewidth of our $^{28}$Si sample and below that of our CZ SOI samples. Among these measurements, however, this comparison between FZ $^{\text{nat}}$Si samples provides the best comparison between implanted and irradiated linewidths as each used similar material outside the ``semiconductor vacuum'' of $^{28}$Si~\cite{saeedi2013}. We thus report only a $\sim35\%$ increase in the inhomogeneous broadening for our implanted ensembles compared to ensembles generated by electron irradiation, and we expect that the broader lines we observe in CZ SOI will be reduced for \textit{T} centres created by implantation into a FZ SOI device layer.

\section{Hyperpolarization}

In other solid state emitters that lack inversion symmetry, spectral diffusion has been shown to increase when centres are generated by ion implantation~\cite{vandam2019, kasperczyk2020} or as centres approach interfaces due to a dramatic increase in electric and magnetic field noise in these regions~\cite{ishikawa2012,evans2016, crook2020}. Typically, spectral diffusion is measured by tracking spectral lines of individual emitters. Here, we develop a novel technique for measuring spectral diffusion within an ensemble of emitters and use it to compare the optical properties of the \textit{T} centres in our SOI samples to those in bulk samples. 

\begin{figure}[ht]
\includegraphics[width=\linewidth]{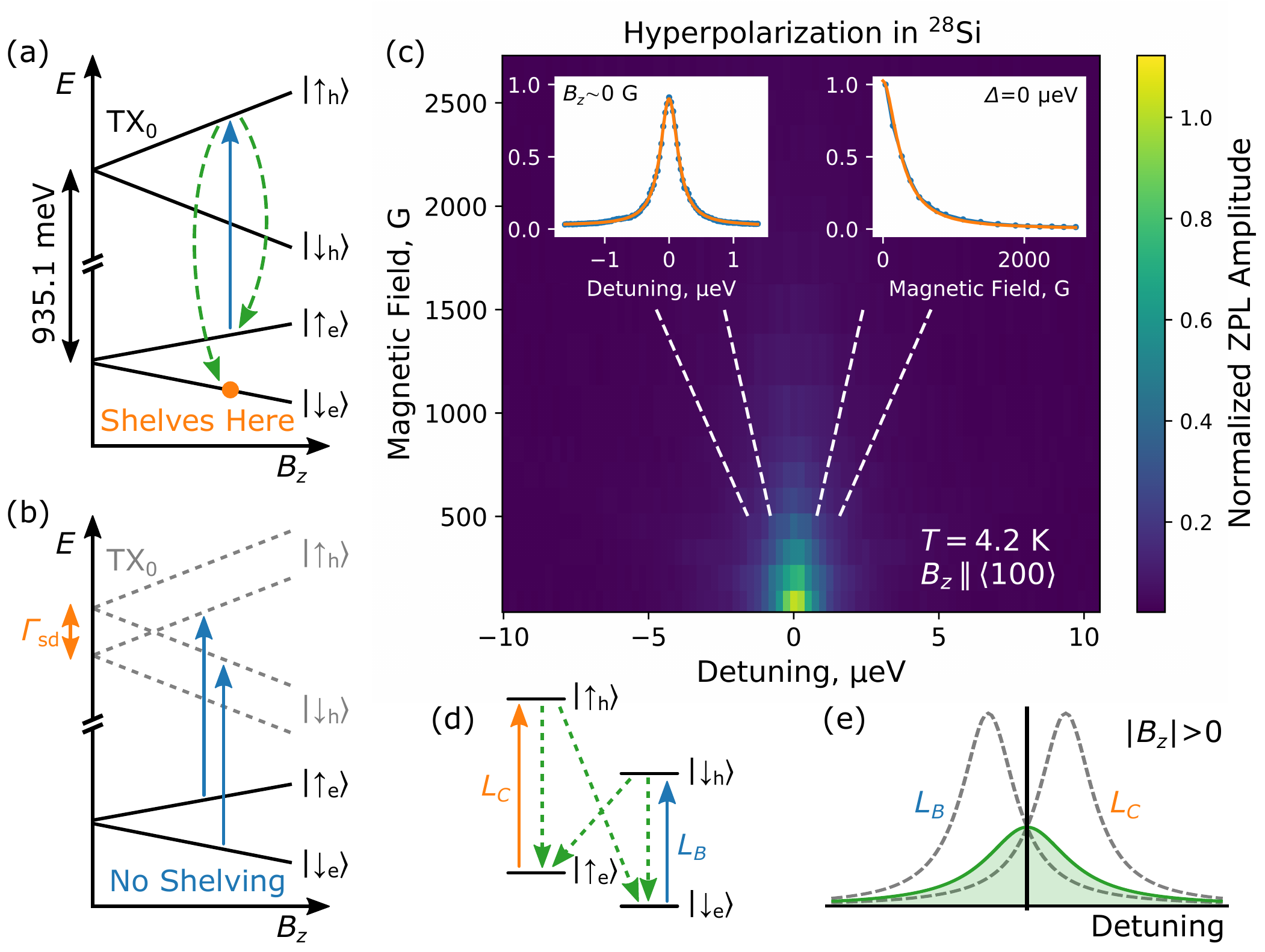} \\
\caption[fig:fig2] {Hyperpolarization in \textit{T} centres (a) Hyperpolarization mechanism for \textit{T} centres. (b) Depiction of how spectral diffusion can limit \textit{T} centre hyperpolarization. (c) PLE measurements of \textit{T} centres in $^{28}$Si performed as a function of magnetic field at a temperature of 4.2~K. Detuning is measured from the ZPL energy of $935.062$~meV. The dashed lines indicate the maximum and minimum Zeeman splittings expected for this magnetic field and sample orientation. The insets show a Lorentzian fit to the residual-field PLE spectrum and a fit of the $\Delta=0$~\textmu eV linecut to the hyperpolarization model developed in the text. (d) Levels and transitions included in the rate model for spin-dependent optical transitions. $L_{\text{B(C)}}$ represents optical excitation in the spin-down (spin-up) manifold, and the green arrows show fluorescent relaxation pathways. (e) Schematic depiction of the loss in PLE signal as a magnetic field splits the \textit{T} centre spin sublevels. The resulting PL spectrum (green curve) is described by Eq.~\ref{eq:model}. }
\label{fig:fig2}
\end{figure}

Under spin-selective resonant optical driving, the \textit{T} centre ground state electron spin will hyperpolarize through a process shown schematically in Fig.~\ref{fig:fig2}a. This unpaired electron provides a spin degree of freedom with an isotropic $g$-factor of $g_e=2.005$ in the orbital ground state. Upon resonant generation of a bound exciton (BE), the ground state electron pairs with the BE electron to form a spin-$0$ singlet. The BE hole then provides an unpaired spin with an anisotropic $g$-factor $g_h$ in the BE $\text{TX}_0$ state. A magnetic field $B_z$ thus generates a spin splitting $g_e\mu_BB_z$ ($g_h\mu_BB_z$) in the ground (excited, or $\text{TX}_0$) state where $\mu_B$ is the Bohr magneton. Because these splittings can differ in magnitude, for Zeeman splittings $\varepsilon_{B}(B_z)=(\pm g_e\pm g_h)\mu_B B_z$ larger than the homogeneous optical linewidth, only one spin sublevel will be excited by the resonant optical drive. Spin non-conserving relaxation will then shelve the system in whichever state is not addressed by the optical drive, hyperpolarizing the electron spin as the \textit{T} centre is optically cycled. For the \textit{T} centre, this hyperpolarization mechanism is efficient~\cite{bergeron2020}.

Using this process, we can reveal the characteristic homogeneous optical linewidth of an ensemble by measuring how the resonantly-driven ZPL evolves in a magnetic field. In the absence of any homogeneous broadening effects, an individual \textit{T} centre's optical linewidth will approach its lifetime-limited lower bound and observable hyperpolarization at low magnetic fields is expected. This onset of hyperpolarization only weakly depends on the inhomogeneously-broadened ensemble lineshape and will apply to ensembles in $^{28}$Si and $^{\text{nat}}$Si alike. Homogeneous effects such as thermal broadening or spectral diffusion will delay this onset of hyperpolarization to larger magnetic fields as shown schematically in Fig.~\ref{fig:fig2}b. If these homogeneous mechanisms have a characteristic broadening on the same scale as the spin splitting, they will ensure that each spin sublevel is intermittently addressed by the optical field, reducing the level of hyperpolarization in the system and increasing the measured PL. 

In Fig.~\ref{fig:fig2}c, we perform PLE spectroscopy of the \textit{T} centre ZPL in the bulk $^{28}$Si sample by scanning a tunable laser over the optical resonance and measuring the PL emission within the collection window highlighted in Fig.~\ref{fig:fig1}d. Repeating this measurement as a function of $B_z$, we see that, rather than Zeeman splitting along the dashed lines of Fig.~\ref{fig:fig2}c, the PLE trace remains stationary and slowly decays as the interrogated \textit{T} centres hyperpolarize. This measurement was performed at a temperature of $4.2$~K where the ZPL linewidth is thermally-broadened~\cite{bergeron2020} and with $B_z$ aligned along the $\langle100\rangle$ crystal axis. For this field orientation, we expect the hole $g$-factors for the twelve possible \textit{T} centre orientational subsets relative to $B_{\langle100\rangle}$ to take on the values $g_h=\{0.91, 2.55\}$ with multiplicities of $4$ and $8$, respectively~\cite{safanov1995, SI}. 

The measured hyperpolarization dynamics can be modeled by noting that both spin sublevels must be driven to prevent population shelving. With a magnetic field applied, the spin sublevels split by $\varepsilon_{B}(B_z)$ where each sublevel is assumed to have the same optical homogeneous linewidth $\Gamma$. As the spin states separate and the probability of driving both sublevels decreases, the system starts to hyperpolarize. Assuming weak optical driving, we model this process with a rate model using the levels and transitions shown schematically in Fig.~\ref{fig:fig2}d and find that the amplitude of a partially hyperpolarized PLE spectrum (Fig.~\ref{fig:fig2}e) is proportional to 
\begin{equation}
A(B_z, \Delta)\propto \frac{L_{\text{B}}(B_z, \Delta)L_{\text{C}}(B_z, \Delta)}{L_{\text{B}}(B_z, \Delta)+L_{\text{C}}(B_z, \Delta)}
\label{eq:model}
\end{equation}
where $L_{\text{B}(\text{C})}(B_z, \Delta)$ is the Lorentzian amplitude of the spin-down (spin-up) sublevel and $\Delta$ is the optical detuning from the zero-field resonance. For simplicity, we have neglected optical transitions between states with different spin orientations, which have smaller spectral overlap with other transitions and have been shown for \textit{T} centres to be much weaker than optical transitions between spin states of the same orientation~\cite{bergeron2020}. The effect of these transitions is examined in the SM~\cite{SI}. 

Evaluating Eq.~\ref{eq:model} and normalizing to the $B_z=0$, $\Delta=0$ value, we obtain an expression for the PLE amplitude as a function of magnetic field for a single \textit{T} centre:
\begin{equation}
A_{\text{h}}(B_z, \Delta)=\frac{\Gamma^2}{4\Delta^2+\Gamma^{2}+\varepsilon_{B}(B_z)^2}.
\label{eq:single}
\end{equation}
Because we simultaneously interrogate a large number of \textit{T} centres in twelve possible orientational subsets, we model our measurement by averaging Eq.~\ref{eq:single} across the twelve values of $g_h$. Here, we have assumed that each orientation contributes equally to the measured signal. 

Fitting the data in Fig.~\ref{fig:fig2}c to this model, we extract a homogeneous linewidth of $\Gamma=$~\chaiSDhot, which agrees very well with the purely thermal linewidth of $230$~MHz at $4.2$~K calculated from the thermal broadening data in Ref.~\cite{bergeron2020} as well as the $\Lambda_{\text{PLE}}=~$\chaiLWhot~linewidth obtained by fitting the PLE spectrum in the $B_{\langle100\rangle}=36$~G residual magnetic field of our setup. These fits are shown in the insets to Fig.~\ref{fig:fig2}c, and the SM provides additional plots demonstrating good agreement between the modeled and measured data~\cite{SI}. The large uncertainty in $\Gamma$ is dominated by uncertainties in sample alignment and $g$-factor calculations as detailed in the SM~\cite{SI}. Simultaneous magnetic resonance measurements in a magnetic field could address these sources of uncertainty by providing \textit{in situ} calibration of the $g$-factors for an arbitrary sample orientation.

\section{Spectral Diffusion in Bulk Si}

\begin{figure}[ht]
\includegraphics[width=\linewidth]{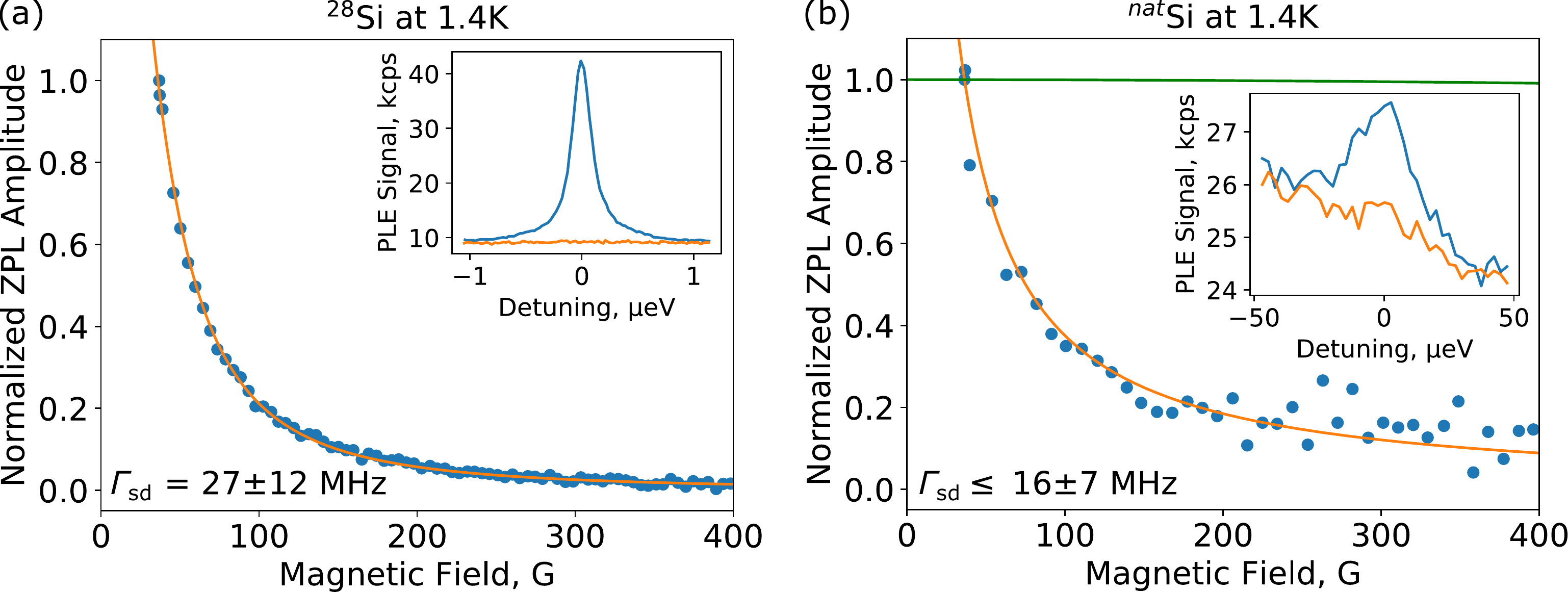} \\
\caption[fig:fig3] {Spectral diffusion in bulk silicon (a,b) \textit{T} centre zero phonon line (ZPL) amplitude as a function of magnetic field at a temperature of 1.4~K for (a) $^{28}$Si and (b) $^{\text{nat}}$Si. The data in (a) is fit to the homogeneously-broadened model given by Eq.~\ref{eq:single}. The model curves in (b) assume dynamics dictated either by $\Gamma_{\text{sd}}$ with inhomogeneous contributions (orange line) or by the optically-measured inhomogeneous linewidth (green line). The insets in (a,b) show the \textit{T} centre ZPL measured at $B_{\langle100\rangle}\sim36$~G (blue) and $\sim450$~G (orange). In the insets to (a) and (b), detuning is measured from the ZPL resonance at $935.064$~meV and $935.167$~meV, respectively.}
\label{fig:fig3}
\end{figure}

When cooled to $1.4$~K, the thermal broadening contribution to the ZPL linewidth is $33$~kHz, and the ZPL PLE linewidth in the $^{28}$Si sample narrows to \makebox{$\Lambda_{\text{PLE}}=$~\chaiLW~\cite{SI}}. In the absence of significant thermal broadening, we may assume that measuring $\Gamma$ amounts to measuring the total spectral diffusion linewidth $\Gamma_{\text{sd}}$. To measure $\Gamma_{\text{sd}}$ at $1.4$~K, we resonantly drive the \textit{T} centre ZPL and ramp $B_{\langle100\rangle}$. Fitting the normalized data shown in Fig.~\ref{fig:fig3}a to Eq.~\ref{eq:single} averaged over the twelve orientational subsets, we find $\Gamma_{\text{sd}}=$~\chaiSD~for our bulk $^{28}$Si sample which agrees with the measured $\Lambda_{\text{PLE}}$. We stress this model only considers homogeneous broadening and ignores any inhomogeneities that might be present. For our measurements of $\Gamma_{\text{sd}}$, we integrate PL counts for $\geq1$~s at each value of $B_{\langle100\rangle}$, making this a long-term average value of the characteristic spectral diffusion within the \textit{T} centre ensemble. 

To demonstrate that this method can provide a means of measuring ensemble homogeneous linewidths that are not resolved in an inhomogeneously broadened PLE spectrum, we repeat this measurement on \textit{T} centres in our bulk FZ $^{\text{nat}}$Si sample, which has an inhomogeneously-broadened optical linewidth of $\Lambda_{\text{PLE}}=$~\taurusLW~at $1.4$~K. To include this inhomogeneous broadening in our hyperpolarization model, we replace $\Delta\rightarrow\Delta + \delta E$ in Eq.~\ref{eq:single} to account for inhomogeneously-shifted resonances and convolve the result with a distribution of $\delta E$ described by a Gauss-Lorentz product with a full width at half maximum equal to the measured optical linewidth $\Lambda_{\text{PLE}}$. For simplicity, we restrict ourselves to $\Delta=0$, and once again, we model contributions from different \textit{T} centre orientational subsets by averaging the calculated response over the expected distribution of $g_h$.


Because our single crystal $^{\text{nat}}$Si sample has been cut along an unknown crystallographic axis, we can only extract an upper bound on $\Gamma_{\text{sd}}$ by fitting the measured data shown in Fig.~\ref{fig:fig3}b as a function of magnetic field alignment and selecting the orientation that maximizes $\Gamma_{\text{sd}}$ as detailed in the SM~\cite{SI}. From this maximal orientation, we find $\Gamma_{\text{sd}}\leq$~\taurusSD~(the orange curve in Fig.~\ref{fig:fig3}b). For comparison, we then assume the measured decay is entirely dictated by inhomogeneous broadening and plot the behaviour predicted by Eq.~\ref{eq:single} with $\Gamma$ equal to the inhomogeneous linewidth $\Lambda_{\text{PLE}}=$~\taurusLWnoPM. The resulting model curve (shown in green in Fig.~\ref{fig:fig3}b) fails to match the data. As these models are calculated for bounding cases, we can comfortably conclude that this method extracts linewidths narrower than $\Lambda_{\text{PLE}}$.

\section{Spectral Diffusion in SOI}

Lastly, we use this method to benchmark the optical properties of implanted \textit{T} centres in CZ SOI. At $1.4$~K, the SOI samples shown in Fig.~\ref{fig:fig1}d display inhomogeneously-broadened ZPL linewidths of $\Lambda_{\text{PLE}}=$~\qqLW~and~\aaLW~for the samples implanted with fluences of $7\times10^{12}$~cm$^{-2}$ and $7\times10^{13}$~cm$^{-2}$, respectively. We measure the PLE amplitude as a function of $B_{\langle100\rangle}$ over the range of $B_{\langle100\rangle}$ for which our optical setup remains stable, and we find $\Gamma_{\text{sd}}=$~\qqSD~and~\aaSD, respectively. Notably, the sample that received the lower implant fluence shows a slightly reduced level of spectral diffusion. This suggests that at least some portion of $\Gamma_{\text{sd}}$ could be attributable to our fabrication process and might be improved with additional surface treatment~\cite{sangtawesin2019}, reduced implant fluence~\cite{vandam2019, wolfowicz2020}, or further processing. 

\begin{figure}[ht]
\includegraphics[width=\linewidth]{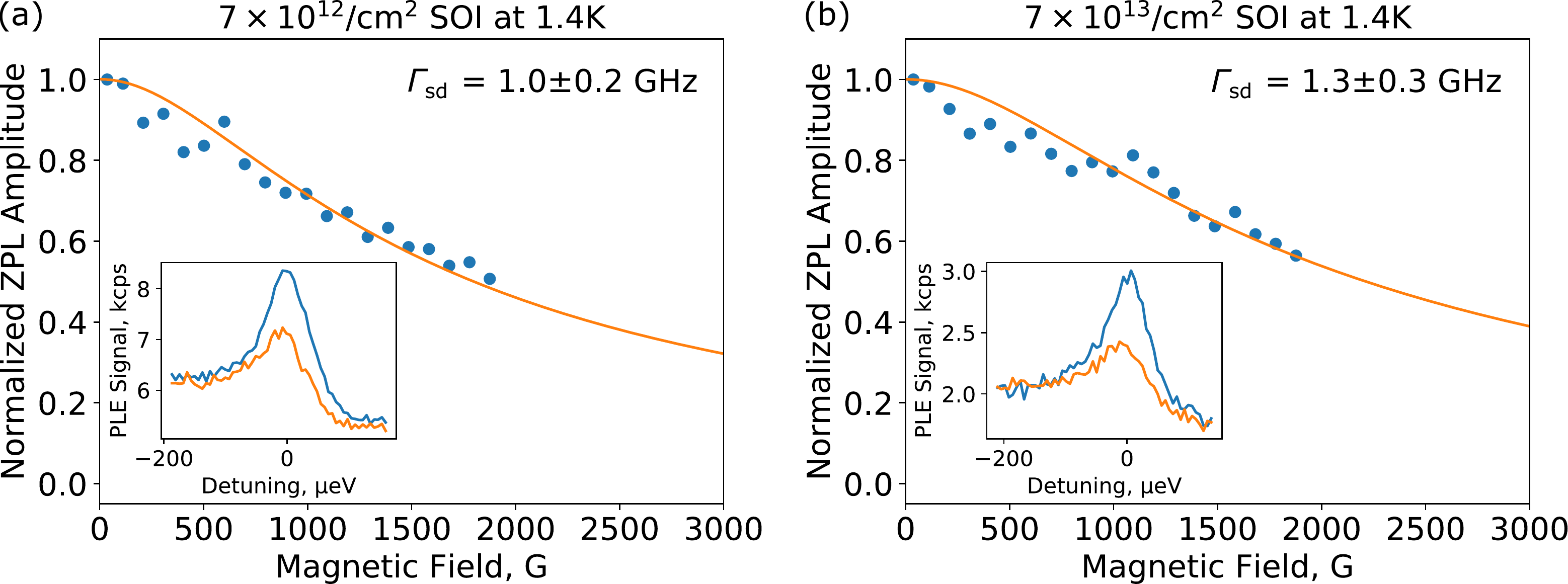} \\
\caption[fig:fig4] {Spectral diffusion in silicon-on-insulator (SOI) (a,b) \textit{T} centre zero phonon line (ZPL) amplitude as a function of magnetic field at a temperature of 1.4~K for SOI implanted at a fluence of (a) $7\times10^{12}$~cm$^{-2}$ and (b) $7\times10^{13}$~cm$^{-2}$. The insets show the \textit{T} centre ZPL measured at $B_{\langle100\rangle}\sim36$~G (blue) and $\sim2000$~G (orange). In the insets to (a) and (b), detuning is measured from the ZPL resonance at $935.176$~meV and $935.237$~meV, respectively. }
\label{fig:fig4}
\end{figure}

It is important to note that the assumption of equally weighted contributions from each \textit{T} centre orientation is likely not true. To address this, we have also fit our data under the assumption that the measured signal comes entirely from the \textit{T} centre orientation that minimizes (maximizes) $\Gamma_{\text{sd}}$ to obtain the bounding sets $\Gamma_{\text{sd}}\in[0.8\pm0.3, 1.7\pm0.1]$~GHz and $\Gamma_{\text{sd}}\in[1.0\pm0.3, 2.1\pm0.1]$~GHz for the $7\times10^{12}$~cm$^{-2}$ and $7\times10^{13}$~cm$^{-2}$ samples, respectively. Once again, simultaneous magnetic resonance measurements in a magnetic field could improve the precision of this technique by calibrating the relative signal contributions of different orientations. 

Even with this uncertainty, however, this method provides efficient comparative measurements of $\Gamma_{\text{sd}}$ and rapid feedback on fabrication processes designed to reduce $\Gamma_{\text{sd}}$. Measuring PLE decay due to spin-dependent optical transitions can provide the characteristic $\Gamma_{\text{sd}}$ within an ensemble of emitters when complementary techniques such as spectral hole burning~\cite{volker1989} are not possible. Moreover, because our measurements interrogate a large ensemble of \textit{T} centres simultaneously, the technique developed here enables quick access to the characteristic value of $\Gamma_{\text{sd}}$ across a sample, which can be time consuming to obtain by measuring individual centres.

\section{Outlook}

Although far from the lifetime-limited linewidth of $\Gamma_1=169$~kHz~\cite{bergeron2020} and $\sim100$ times larger than the values we report in bulk samples, these values for $\Gamma_{\text{sd}}$ of implanted \textit{T} centres in CZ SOI are already competitive with other solid state emitters near interfaces. It is common for spectral diffusion in defect centres that lack inversion symmetry to be on the order of several GHz in the absence of careful surface treatment or experimental protocols~\cite{ishikawa2012, evans2016, anderson2019, vandam2019, crook2020}. With minimal optimization, implanted \textit{T} centres display $\sim1$~GHz characteristic spectral diffusion, and this could be minimized further as fabrication and measurement techniques continue to develop. Moreover, the characteristic spectral diffusion reported here is a long-term, ensemble average. Individual centres within this ensemble could have much lower $\Gamma_{\text{sd}}$~\cite{chu2011, kasperczyk2020}, and any slowly varying effects that contribute to this long-term average might be addressed by feedback or filtering techniques~\cite{bernien2013, gao2019}.

To estimate the effect spectral diffusion will have on the performance of a \textit{T} centre-based photon-spin interface, we calculate the indistinguishability of two photons emitted by a given \textit{T} centre as $I=\xi\Gamma_{1}/(\xi\Gamma_{1}+\Gamma_{\text{sd}})$ where $\xi=0.23$ is the \textit{T} centre Debye-Waller factor~\cite{bylander2003,grange2015, bergeron2020}. For our bulk and SOI samples, respectively, we find $I\sim2\times10^{-3}$ and $I\sim4\times10^{-5}$. We stress that both of these values can likely be improved by employing fabrication or measurement techniques designed to stabilize the electromagnetic environment~\cite{oliveira2017, anderson2019, sangtawesin2019, kasperczyk2020} or by filtering the emitted photons~\cite{bernien2013, gao2019}. Moreover, embedding \textit{T} centres inside optical resonators can Purcell enhance $\Gamma_1$ to increase the photon indistinguishabilty. Optical resonators in SOI capable of demonstrating Purcell factors above $10^4$ have already been demonstrated~\cite{zain2008}. Because the \textit{T} centre radiative efficiency is expected to be near-unity~\cite{bergeron2020}, \textit{T} centres incorporated into such an optical cavity could provide high quality photon-spin interfaces in the telecommunications $O$-band. 

In summary, we have demonstrated a reliable method for using ion implantation and annealing treatments to incorporate \textit{T} centres into industry-standard silicon integrated photonics material. We then developed an efficient method, which could be applied to a wide variety of colour centres, for using spin-dependent optical transitions to benchmark spectral diffusion. Finally, we used this new technique to show that implanted \textit{T} centres within a $220$~nm SOI device layer can display GHz-scale levels of spectral diffusion with minimal optimization. These results represent an important step towards developing quantum technologies in silicon. 

%

\section{Acknowledgments}

We thank M. L. W. Thewalt and L. Childress for fruitful discussions as well as N. V. Abrosimov for bulk sample growth and C. Cl\'{e}ment for rapid thermal annealing. 

\section{Funding}

This work was supported by the Natural Sciences and Engineering Research Council of Canada (NSERC), the Canada Research Chairs program (CRC), the Canada Foundation for Innovation (CFI), the B.C. Knowledge Development Fund (BCKDF), the Canadian Institute for Advanced Research (CIFAR) Quantum Information Science program, the CIFAR Catalyst Fund, and Le Fonds de recherche du Qu\'{e}bec – Nature et technologies (FRQNT). The $^{28}$Si samples used in this study were prepared from the Avo28 crystal produced by the International Avogadro Coordination (IAC) Project (2004--2011) in cooperation among the BIPM, the INRIM (Italy), the IRMM (EU), the NMIA (Australia), the NMIJ (Japan), the NPL (UK), and the PTB (Germany).

\section{References}
\bibliography{megaBib}

\section{Supplementary Information}

\subsection{Measurement Details}

The samples are mounted in a liquid helium cryostat. The sample temperature is set by pumping on the liquid helium bath. The PL and PLE spectroscopy measurements were performed as reported in Ref.~\cite{bergeron2020}. For PLE spectroscopy measurements a Toptica DL100 tunable diode laser was used for measurements of $^{28}$Si, and a tunable nanoplus DFB laser diode was used for the broader SOI and bulk $^{\text{nat}}$Si samples. 

\subsection{Implantation Dose Optimization}

To select an implant dose for our \textit{T} centre implantation study, we first generated \textit{T} centres in float zone (FZ) silicon wafers by co-implanting carbon and hydrogen. We swept the fluence of the implants while keeping fixed C:H ratios of 1:1, 2:1, and 1:0. Carbon implants were again performed at an energy of $38$~keV, and hydrogen was implanted at $9$~keV. The samples were then annealed in air following a $(100, 300, 350, 400, 450)^{\circ}$C step-wise temperature profile with dwell times of $(60, 30, 30, 60, 60)$~min. This temperature profile was chosen as it previously generated high density \textit{T} centre concentrations in electron-irradiated, high-carbon bulk samples. 

\begin{figure}[ht]
\includegraphics[width=\linewidth]{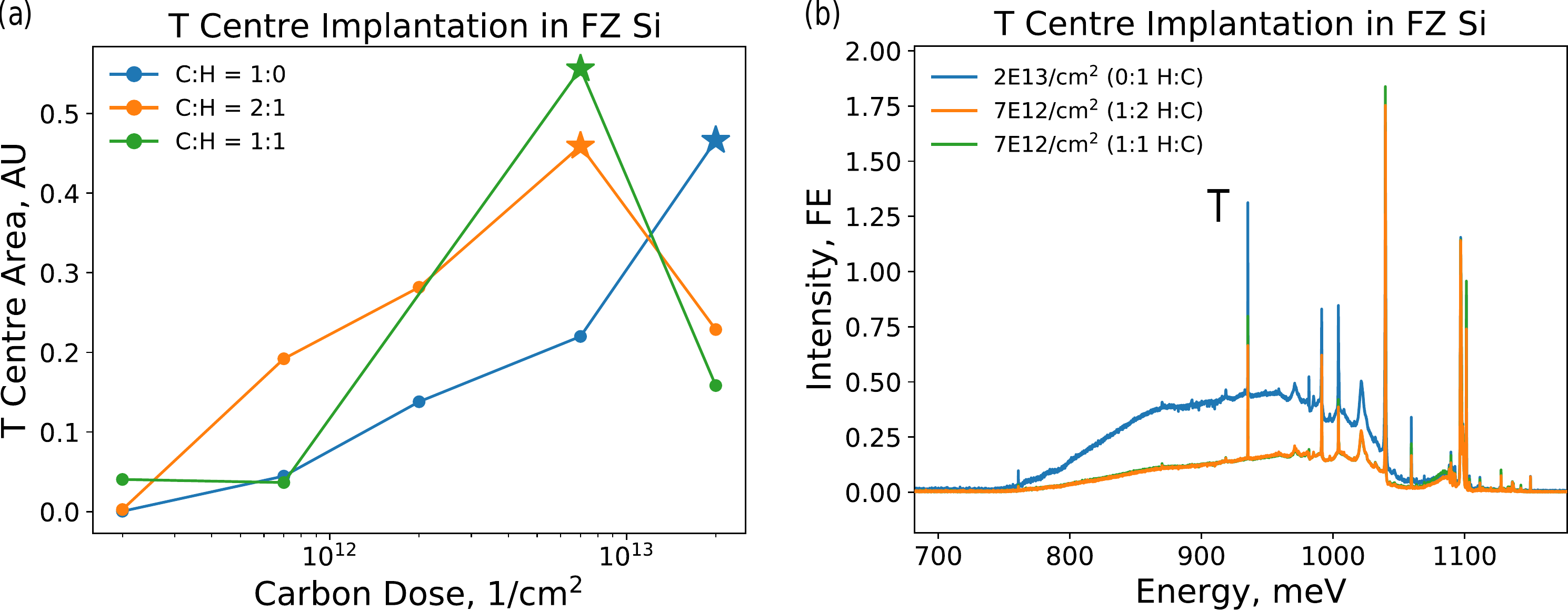} \\
\caption[fig:figSdose] {\textit{T} centre implantation development. (a) Integrated \textit{T} centre zero phonon line (ZPL) area as a function of implant dose for fixed carbon-to-hydrogen ratios. (b) \textit{T} centre photoluminescence (PL) spectra for carbon and hydrogen co-implantation into bulk float zone silicon.}
\label{fig:figSdose}
\end{figure}

Fig.~\ref{fig:figSdose}a shows the integrated area of the FE-normalized \textit{T} centre ZPL for each fluence we measured, and Fig.~\ref{fig:figSdose}b presents the PL spectra for the implant fluences that generated the brightest \textit{T} centre ZPL for each C:H ratio. Each spectrum shows a strong \textit{T} centre ZPL, but the 1:0 C:H sample has a much higher broadband radiation damage background. By comparison, the 1:1 and 2:1 C:H implants show comparable \textit{T} centre intensity with a much lower background for carbon fluences of $7\times10^{12}$~cm$^{-2}$. From this data, we decided to fix the C:H ratio at 1:1 and implant at fluences of $7\times10^{12}$~cm$^{-2}$ and $7\times10^{13}$~cm$^{-2}$.


\subsection{T Centre Localization}

\begin{figure}[ht]
\includegraphics[width=\linewidth]{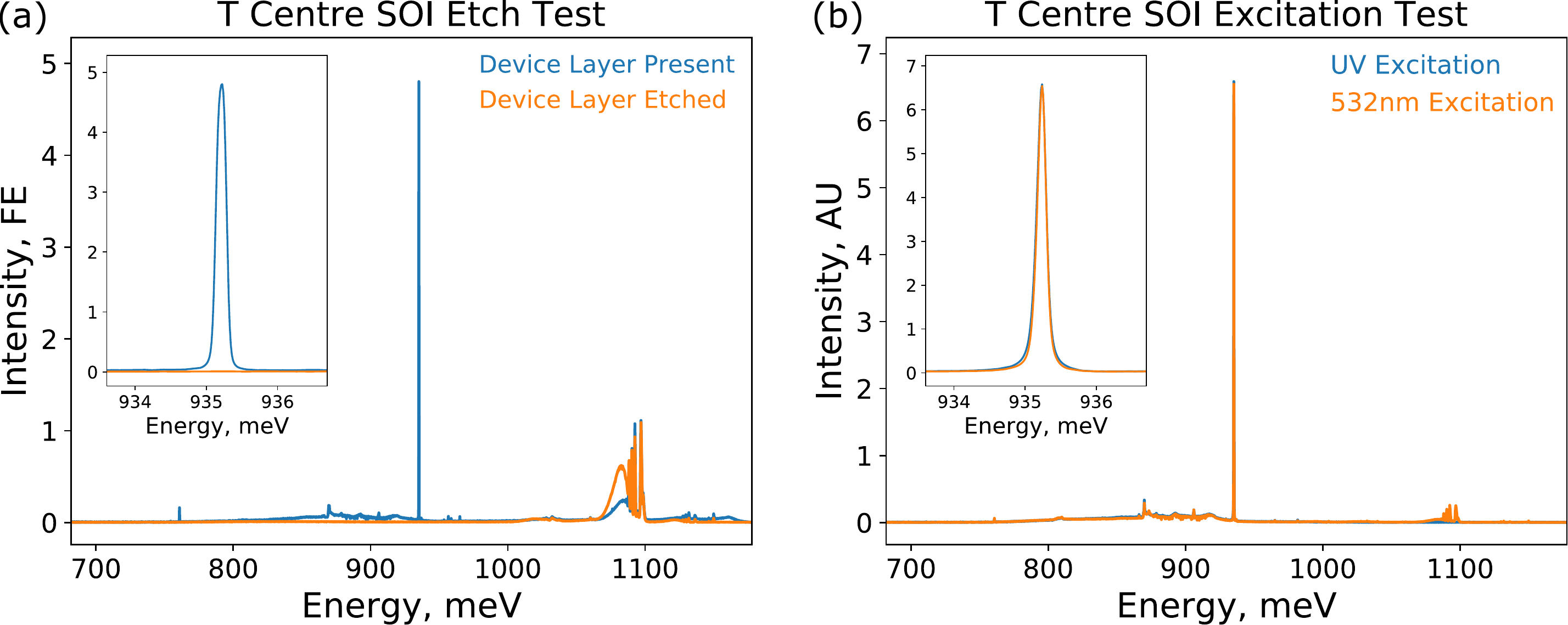} \\
\caption[fig:uvPL] {\textit{T} centre localization. (a) Photoluminescence (PL) spectra of an implanted sample that has had a portion of its device layer etched away. (b) PL spectra of implanted SOI samples under $532$~nm and UV excitation. }
\label{fig:uvPL}
\end{figure}

To confirm the generated \textit{T} centres are localized within the SOI device layer, we etched away the device layer from one of our SOI samples and measured PL from the exposed buried oxide and handle wafer material. As shown in Fig.~\ref{fig:uvPL}a, the resulting $532$~nm PL spectrum shows no significant \textit{T} centre PL as compared to the same sample in an unetched region, suggesting that the handle material does not contain a significant concentration of \textit{T} centres. 

As further confirmation that the generated \textit{T} centres are localized within the SOI device layer, we performed PL spectroscopy with UV excitation. Because the absorption depth of $532$~nm light in Si is $2.5$~\textmu m at cryogenic temperatures, the SOI PL spectra presented in the main text sample both the SOI device layer and the silicon handle wafer beneath the buried oxide. UV wavelengths have a penetration depth of $<1$~\textmu m in Si at cryogenic temperatures~\cite{dash1955, jellison2018}, making PL with UV excitation more sensitive to the device layer than PL with $532$~nm excitation. Fig.~\ref{fig:uvPL}b shows a representative PL spectrum measured on the same sample with both $532$~nm excitation and UV excitation. Because the UV measurement shows a negligible FE recombination line, we have normalized both spectra to their integrated signal. Both spectra show strong \textit{T} centre lines, confirming that there are \textit{T} centres within the device layer. 

Taken together, these two measurements localize our implanted \textit{T} centres to within the device layer of the SOI materal. 

\subsection{Calculating g-factors}

Our calculation of the hole $g$-factors expands upon the work of Ref.~\cite{safanov1995}. We write the hole Hamiltonian as the sum of a strain and a magnetic term $H_h = H_s(\epsilon)+H_b(B)$ where the strain Hamiltonian is
\begin{equation}
H_s(\epsilon)=-b\sum_i (J_i^2-I)\epsilon_{ii} - \frac{d}{\sqrt{3}}\sum_{i\neq j}[J_i,J_j]\epsilon_{ij}.
\end{equation}
Here, $J_i$ are the angular momentum operators, $[J_i,J_j]=\frac{1}{2}(J_iJ_j+J_jJ_i)$, $\epsilon_{ij}$ are strain tensor components, and $b$, $d$ are deformation parameters. An external magnetic field $\vec{B}$ enters the Hamiltonian as
\begin{equation}
H_b(\vec{B}) = \mu_B\left(g_1\sum_i B_i J_i + g_2\sum_i B_i J_i^3\right).
\end{equation}
where $\mu_B$ is the Bohr magneton and $g_1$, $g_2$ are the hole $g$-factors. 

We use the internal strain and deformation parameters obtained by Ref.~\cite{safanov1995}: $\epsilon_{YY}=-0.65\times10^{-3}$, $\epsilon_{ZZ}=-0.26\times10^{-3}$, $b=-0.8$~eV, and $d=-2.7$~eV. The internal strain parameters were defined relative to the defect axes where $\hat{y}$ is parallel to $\langle110\rangle$ and $\hat{z}$ is in the $(110)$ plane, tilted from the $\langle001\rangle$ axis by $4^{\circ}$. In our treatment, we rotate the internal strain from the defect coordinate system to that of the crystal so that magnetic field alignments can be referenced to crystallographic axes. 

Rather than using the $g_1=1.3$ and $g_2=-0.1$ values from Ref.~\cite{safanov1995}, we obtain $g_1=1.505\pm0.022$ and $g_2=-0.138\pm0.012$ by fitting the model above to the $g$-factors obtained in Ref.~\cite{bergeron2020}, restricting ourselves to $g_i$ values near those reported in Ref.~\cite{safanov1995}. The highly-resolved $^{28}$Si data from Ref.~\cite{bergeron2020} enables us to fit the hole $g$-factors using all twelve \textit{T} centre orientational subsets. For this fit, we also introduce a misalignment of the magnetic field as an additional free parameter. Our fit returns an angular inclination of $10.1^{\circ}\pm0.1^{\circ}$ and rotation of $71.9^{\circ}\pm0.6^{\circ}$ about the nominal $\langle110\rangle$ field alignment in that work. Table~\ref{tab:gFact} provides comparisons between our fit results and the $g_h$ values reported in Ref.~\cite{bergeron2020} for a $\langle110\rangle$ magnetic field alignment.

\begin{table}[ht]
\begin{center}
 \begin{tabular}{|c | c | c | c |} 
 \hline
 Subset & Measured $g_h^{\langle110\rangle}$~\cite{bergeron2020} & Fitted $g_h^{\langle110\rangle}$ & Calculated $g_h^{\langle100\rangle}$ \\ 
 \hline
 1 & $3.457\pm0.007$ & $3.460\pm0.084$ & $0.91\pm0.10$ \\ 
 2 & $3.457\pm0.007$ & $3.459\pm0.084$ & $0.91\pm0.11$ \\
 3 & $2.233\pm0.009$ & $2.269\pm0.057$ & $0.91\pm0.16$ \\
 4 & $2.165\pm0.014$ & $2.252\pm0.057$ & $0.91\pm0.16$ \\
 5 & $1.970\pm0.012$ & $2.036\pm0.052$ & $2.55\pm0.42$ \\
 6 & $1.871\pm0.022$ & $2.015\pm0.052$ & $2.55\pm0.42$ \\
 7 & $1.851\pm0.014$ & $1.779\pm0.049$ & $2.55\pm0.13$ \\
 8 & $1.770\pm0.008$ & $1.758\pm0.049$ & $2.55\pm0.13$ \\
 9 & $1.596\pm0.006$ & $1.566\pm0.044$ & $2.55\pm0.08$ \\
 10 & $1.497\pm0.011$ & $1.537\pm0.045$ & $2.55\pm0.08$ \\
 11 & $1.082\pm0.007$ & $1.029\pm0.024$ & $2.55\pm0.34$ \\
 12 & $1.069\pm0.007$ & $1.023\pm0.024$ & $2.55\pm0.34$ \\
 \hline
\end{tabular}
\caption[fig:figSdose]{Comparing hole $g$-factors for a $\langle110\rangle$ magnetic field orientation as measured in Ref.~\cite{bergeron2020} to those fitted by the model introduced in Ref.~\cite{safanov1995}.}
\label{tab:gFact}
\end{center}
\end{table}

With model parameters in hand, we then calculate $g_h$ for each \textit{T} centre orientational subset by finding the $B_i$ components for each subset, computing the eigenvalues of $H_h$, and dividing the resulting spin splitting by $\mu_BB$. $H_h$ predicts a negligible zero-field splitting within the $\text{TX}_0$ orbital branch and negligible interaction with the $\text{TX}_1$ orbital at the moderate magnetic fields used in this work, justifying this approach for calculating the $g_h$ values. 

The $g$-factors for $\vec{B}\parallel\langle100\rangle$ used in the main text are listed with their calculated error bars in Table~\ref{tab:gFact}. For these calculations, we have assigned inclination and azimuthal $\vec{B}$-field alignment errors of $\pm10^{\circ}$ each. 

\begin{figure}[ht]
\includegraphics[width=\linewidth]{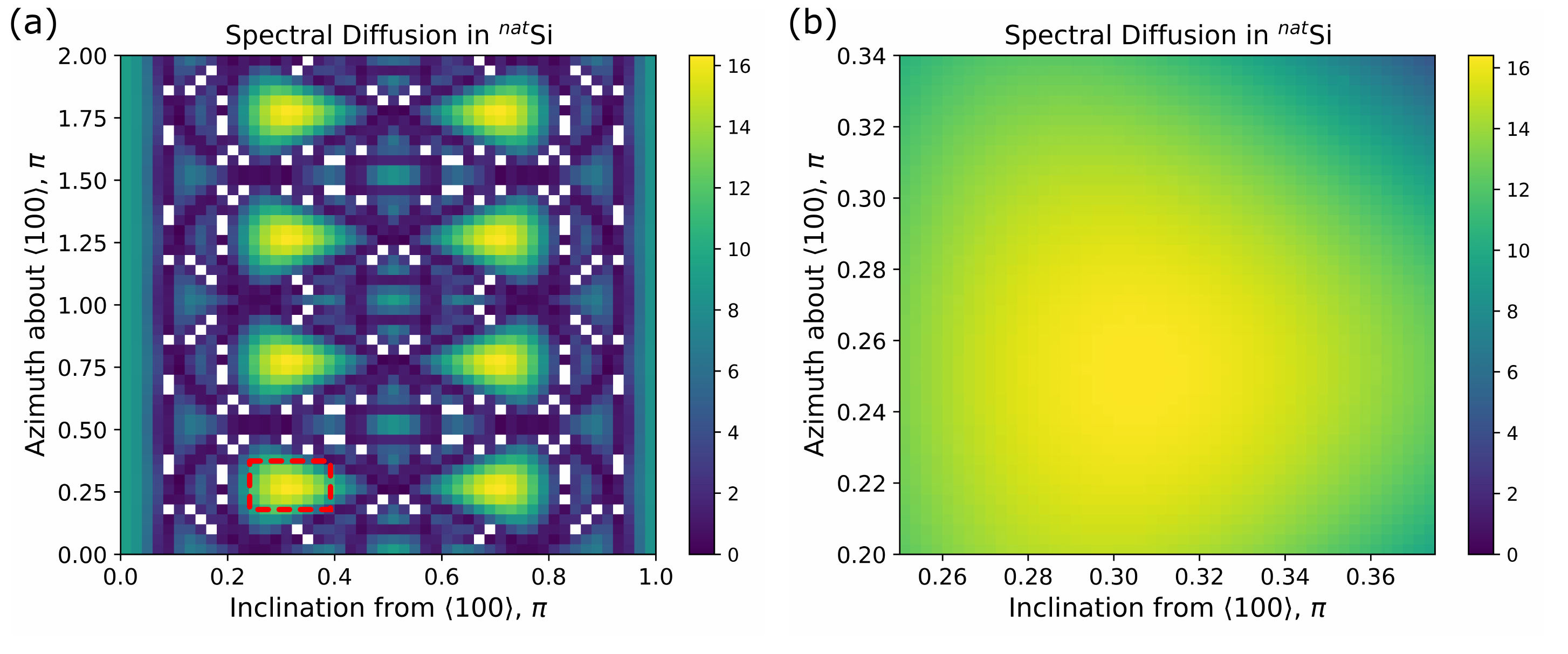} \\
\caption[fig:taurusFits] {Bounding calculation of $\Gamma_{\text{sd}}$ in bulk $^{\text{nat}}$Si. (a,b) Fitted values of $\Gamma_{\text{sd}}$ as a function of magnetic field orientation. The orientations in white were unable to fit the data and were excluded from the plot. (b) A higher resolution sweep of orientations in the red square of (a) that are near the maximal $\Gamma_{\text{sd}}$.}
\label{fig:taurusFits}
\end{figure}

For our bounding calculation of $\Gamma_{\text{sd}}$ in \natSi, we fit the measured data as a function of the magnetic field alignment. The fitted values for $\Gamma_{\text{sd}}$ are plotted as a function of the magnetic field orientation in Fig.~\ref{fig:taurusFits}. Here, we have omitted magnetic field orientations for which the model is unable to fit the measured data. The model curve in Fig.~\ref{fig:fig3}b of the main text is plotted using inclination and azimuthal angles of $0.30\pi$ and $0.25\pi$, respectively. 

\subsection{Optical Transitions Between Different Spin Orientations}

\begin{figure}[ht]
\includegraphics[width=\linewidth]{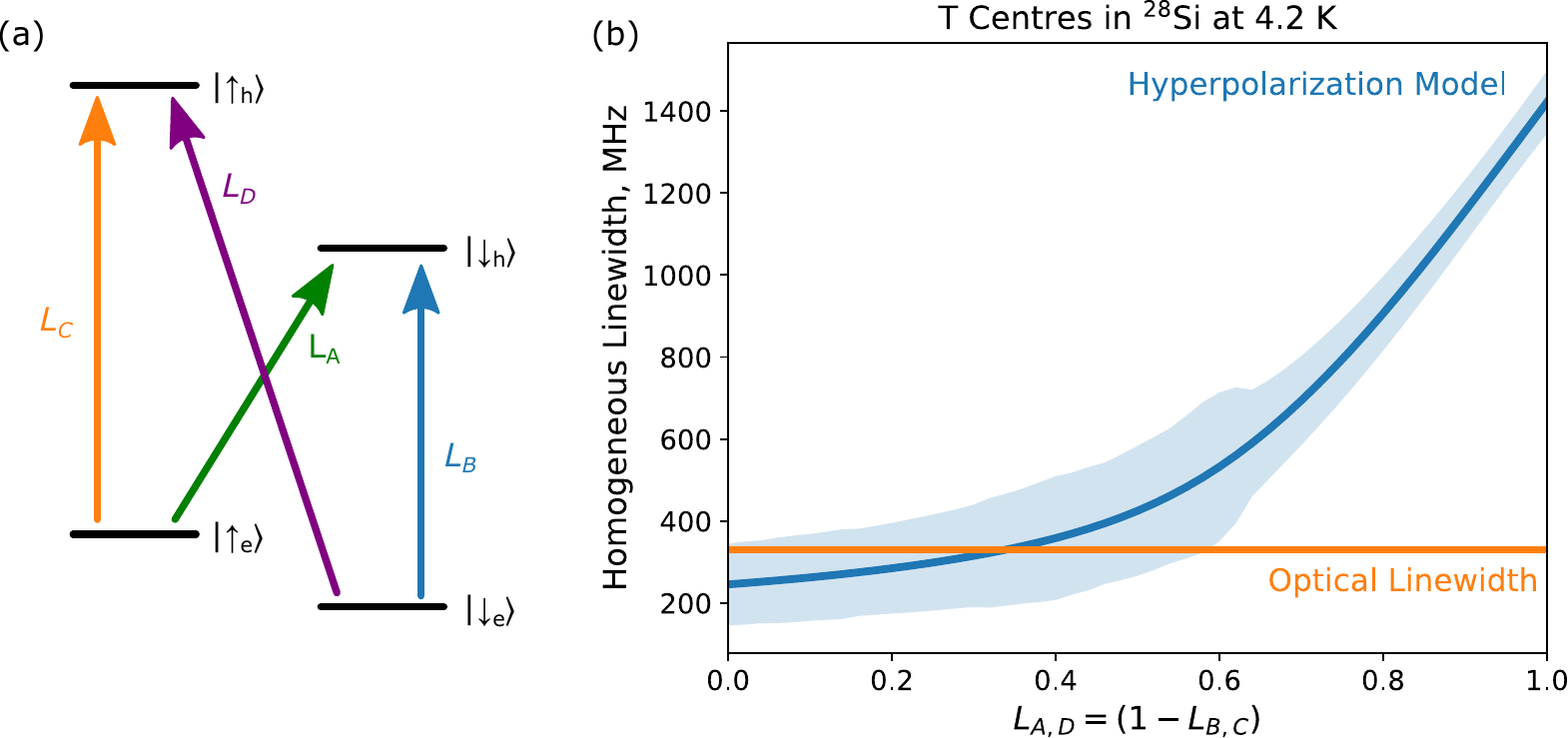} \\
\caption[fig:spinnoncon] {Effect of optical transitions between different spin orientations. (a) \textit{T} centre optical transitions. (b) \textit{T} centre homogeneous linewidth extracted from the $4.2$~K hyperpolarization data presented in the text. Good agreement between the optical and hyperpolarization measurements is only found when transitions between parallel spin states dominate ($L_{B,C}>L_{A,D}$). }
\label{fig:spinnoncon}
\end{figure}

In the model presented in the main text, we have neglected optical transitions between different spin orientations. Including these transitions into the rate model used to derive Eq.~\ref{eq:model} of the main text, we can rewrite the amplitude proportionality as
\begin{equation}
A\propto\frac{L_BL_C + L_AL_C + L_BL_D + L_AL_D}{L_A+L_B+L_C+L_D}
\label{eq:model4}
\end{equation}
where the optical transitions are labeled by subscripts as in Fig.~\ref{fig:spinnoncon}a. The amplitudes $L_B$, $L_C$ have been measured to be $\sim10$ times larger than $L_A$, $L_D$~\cite{bergeron2020, kurkjian2021}. Moveover, since $\left|\frac{\partial L_A}{\partial B_z}\right|,\left|\frac{\partial L_D}{\partial B_z}\right|>\left|\frac{\partial L_B}{\partial B_z}\right|,\left|\frac{\partial L_C}{\partial B_z}\right|$, the approximation given by Eq.~\ref{eq:model} of the main text becomes better as $B_z$ increases.

To quantify the effect of the $L_A$ and $L_D$ optical transitions on our hyperpolarization model, we fit the thermally-broadened $T=4.2$~K data displayed in Fig.~\ref{fig:fig2}c of the main text using the four-transition model of Eq.~\ref{eq:model4}. Ignoring inhomogeneous broadening, we find $\Gamma=250\pm100$~MHz with $L_{A,D}=0$ and $\Gamma=260\pm100$~MHz with $L_{A,D}=0.1L_{B,C}$. To further explore this dependency, we fix $L_{A,D}+L_{B,C}=1$, fit the measured data as a function of $L_{A,D}$, and compare the homogeneous linewidth extracted from the model to the optically measured linewidth $\Lambda_{\text{PLE}}=$~\chaiLWhot. Plotting the results in Fig.~\ref{fig:spinnoncon}b, we see that $\Gamma$ only becomes strongly dependent on $L_{A,D}$ when $L_{A,D}>L_{B,C}$, as expected. In this regime, the hyperpolarization model returns values of $\Gamma$ that are much larger than the optically-measured value, supporting the assertion that $L_{A,D}\ll L_{B,C}$ that has been experimentally verified elsewhere~\cite{bergeron2020, kurkjian2021}.

\subsection{Fitting residual-field PLE spectra}

The poles of the electromagnet used in these measurements generated a residual field of $B_0\sim36$~G. For our $^{28}$Si sample, the Zeeman splitting from this residual field is comparable to the measured optical linewidth $\Lambda_{\text{PLE}}$. To account for this in our analysis, we fit the residual field spectra to an ensemble average of Eq.~\ref{eq:single} with $B_z$ along the $\langle100\rangle$ axis and take $\Gamma=\Lambda_{\text{PLE}}$. 

For the $^{\text{nat}}$Si and SOI samples where the lineshape is better described by the Gauss-Lorentz product (GLP), we define the GLP linewidth to be $\Gamma=\sqrt{\Lambda_{\text{PLE}}^2+\varepsilon_B(B_z)^2}$. 

\subsection{Hyperpolarization model linecuts}

\begin{figure}[ht]
\includegraphics[width=\linewidth]{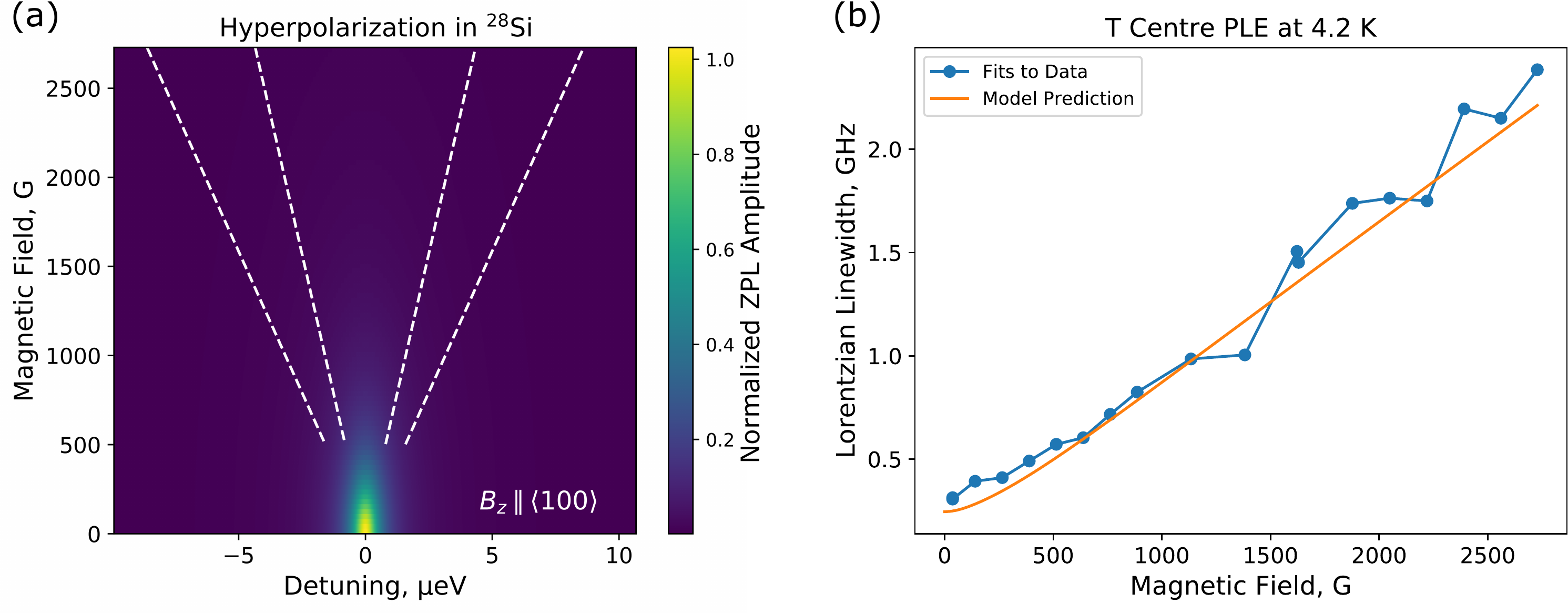} \\
\caption[fig:extraModel] {Hyperpolarization in \textit{T} centres (a) Simulated version of Fig.~\ref{fig:fig2}c in the main text. (b) Predicted and measured PLE linewidth as a function of magnetic field for our $^{28}$Si sample at 4.2~K.}
\label{fig:extraModel}

\end{figure}

In Fig.~\ref{fig:fig2}c of the main text, we measure \textit{T} centre hyperpolarization in our $^{28}$Si sample at $4.2$~K and compare the measured data to the behaviour predicted by the hyperpolarization model developed in the text. As additional comparison, we plot the simulated 2d data in Fig.~\ref{fig:extraModel}a. We then fit both the measured and the modeled PLE spectra to Lorentzian lineshapes and plot the fitted linewidths against each other in Fig.~\ref{fig:extraModel}b. 

\end{document}